\documentclass[prl,aps,twocolumn,floats,showpacs]{revtex4}
\usepackage[dvips]{graphicx}
\usepackage{dcolumn}
\newcommand{\Fe}  {$^{57}$Fe}
\newcommand{\APL} {Appl.Phys.Lett.}
\newcommand{\HI}  {Hyperfine Int.}
\newcommand{\PR}  {Phys.Rev.}
\newcommand{\PRL} {Phys.Rev.Lett.}
\newcommand{\ei}  {\mbox{\scriptsize\sf i}}
\newcommand{\ee}  {\mbox{e}}
\newcommand{\ii}  {\mbox{\sf i}}
\graphicspath{{FiguresLow/}}
\begin{document}
\title{Exo-interferometric Phase Determination of X rays}
\author{Wolfgang Sturhahn}
\email{sturhahn@anl.gov}
\author{Caroline L'abb\'e}
\author{Thomas S.\,Toellner}
\affiliation{Advanced Photon Source, Argonne National Laboratory,
Argonne, Illinois 60439}
\date{\today}
\begin{abstract}
%
The time-dependent phase change of x rays after transmission
through a sample has been determined experimentally.
We used an x-ray interferometer with a reference sample containing
sharp nuclear resonances to prepare a phase-controlled x-ray beam.
A sample with similar resonances was placed in the beam but
outside of the interferometer; hence our use of the term
``exo-interferometric.''
We show that the phase change of the x rays due to the sample can be
uniquely determined from the time-dependent transmitted intensities.
The energy response of the sample was reconstructed from the phase and
the intensity data with a resolution of 23\,neV.
%
\end{abstract}
\maketitle

%
%
Particle scattering, particularly of photons, is one of
the most widely used tools in science.
The measured quantity is usually the particle flux or the intensity.
The phase of the particle field, or more precisely the phase
change of the field during the scattering process, goes undetected
by the intensity measurement.
The related loss of information is known as the ``phase problem''
in various areas of research.

Methods of phase determination are commonly based on
creating two or more indistinguishable paths for the particle
where at least one path does not include the sample.
The resulting interference pattern is then accessible by
intensity measurements.
Interferometric techniques with x rays
\cite{Bonse65,Bonse77,Hirano96},
light \cite{HariharanMalacara}, neutrons \cite{RauchWerner},
or atoms \cite{Berman} have been developed in this context.
Results usually depend on details in the detected interference pattern,
and measurements become exceedingly difficult for short wavelengths.
In the x-ray regime, thermal and mechanical stability of the
interferometer (IFM) are crucial, and the spatial limitations on a
sample in close proximity to sensitive components of the IFM can be
problematic for experimental work.
Here we discuss an x-ray interferometric method that avoids
this problem.

The typical x-ray IFM works with sample and reference
placed in spatially separate paths inside the IFM.
This allows the comparison of phase changes to the
photon field caused by sample and reference.
Besides potential stability problems, interference is not
achieved if the direction of the x rays is changed by a
scattering process in the sample.
The spatial separation of x-ray IFM and sample for phase measurements
would be highly desirable but to our knowledge has not been demonstrated
previously.
In this paper, we present measurements of phase changes
caused by a sample placed \textit{outside} the IFM, i.e.,
after recombination of the beam paths.
We use the term ``exo-interferometry'' to emphasize this fact.
Exo-interferometry permits us to analyze phase changes of x rays
that were transmitted, reflected, or diffracted by a sample material.
Our approach is based on recent theoretical work \cite{Sturhahn01}
on time-dependent phase determination using an x-ray IFM.
For experimental verification, we used a coherent nuclear resonant
scattering process \cite{GerdauDeWaard}, which produces a signal that
is sufficiently delayed to be observable by x-ray detectors.
Previous experiments on nuclear resonant interferometry
\cite{Hasegawa94,Izumi95} placed samples inside the IFM and did not
attempt to obtain the time-dependent phase.
The principles of exo-interferometric phase determination
are elucidated next.

\begin{figure}[t]
\centering
\includegraphics*[scale=0.6]{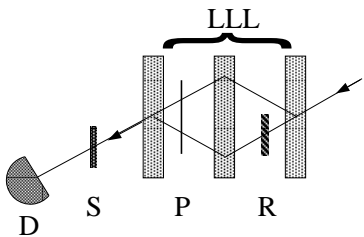}
\caption{Experimental setup.
The incident SR is split at the first face of the triple-Laue
IFM (LLL).
The reference, R, is placed in the lower beam path.
A rotatable 0.5-mm-thick, polished Be platelet works as a phase
retarder, P.
The sample, S, is kept spatially separate from the IFM, except
in the control experiment, where it is placed next to the
reference but in the upper beam path.
The avalanche photodiode detector, D, provides time-resolved
spectra.}
\label{fig:setup}
\end{figure}

The addition of field amplitudes in the IFM allows the production
of a time-dependent field with a phase that can be manipulated in
a controlled way.
Consider a very short x-ray pulse described by $\delta(t)$ that is
incident on a triple-Laue IFM similar to the schematic shown
in Fig.\,\ref{fig:setup}.
If we place a reference material with response function $R(t)$ in one
of the paths of the IFM and apply an additional phase shift $\alpha$
between the two paths, the exit field will be roughly
proportional to $\exp[\ii\alpha]\,\delta(t)+R(t)$.
It will be useful to introduce the delayed response explicitly.
Let $R(t)=r_0\delta(t)+R'(t)$, where $R'(t)$ is the delayed response
that vanishes for ``early'' times $t<\tau$ after arrival of the x-ray pulse.
The field amplitude after the IFM is then given by
\begin{equation}
(r_0+\ee^{\ei\alpha})\,\delta(t) + R'(t)\:\:.
\label{eq:IFMfield}
\end{equation}
At early times, modulus and phase of the field are controlled by
the phase shift $\alpha$, whereas at later times the components of
the field are unaffected.
The basic idea of exo-interferometric phase determination is
now to mix field components with different time coordinates
simply by scattering the x-ray field described by
Eq.\,(\ref{eq:IFMfield}) off a sample.
The degree of mixing will depend on the properties of the
sample and on the adjustable value of $\alpha$.
If the response of the sample material is prompt, i.e., proportional
to $\delta(t)$, mixing does not occur -- a delayed response
is required.
Assume the transmission through a sample with response
function $S(t)=s_0\delta(t)+S'(t)$ with delayed part $S'(t)$.
The expression for the field amplitude after the sample
contains three terms
\begin{equation}
s_0 (r_0+\ee^{\ei\alpha})\,\delta(t) +
\left\{ s_0 R'(t) + (r_0+\ee^{\ei\alpha})\,S'(t) \right\} +
R'{\bf \ast}S'(t)\:\:.
\label{eq:field}
\end{equation}
The first term describes conventional x-ray interferometry.
The second term is reminiscent of the coherent superposition of
delayed reference and sample response that is observed when
placing the sample inside the IFM.
The last term is a convolution integral that also gives a
delayed contribution.
A similar expression, but all terms with $\alpha$ being absent,
has been used to describe ``time-domain interferometry''
\cite{Baron97,Smirnov01}.

The previous arguments lead to the following conclusions about
the feasibility of exo-interferometric experiments.
Reference and sample must produce time-delayed responses, and
the convolution of the responses should be negligible, i.e.,
either the spectra of $R'$ and $S'$ have a small overlap or the
response is sufficiently weak.
There must be a time scale $\tau$ that clearly separates ``early''
and ``delayed.''
However, the time scale does not enter quantitatively -- it
is mostly determined by experimental circumstances.
The response time of an x-ray IFM using single crystals
with low-order Bragg or Laue reflections is of the order 10$^{-15}$\,s.
In the present experiment, the x-ray pulses incident on the IFM had
a longitudinal coherence time of about 0.3\,ps corresponding
to an energy bandwidth of 2\,meV.
In addition, the average length of the synchrotron radiation (SR)
pulse, 70\,ps, and the time resolution of the x-ray detector, 1\,ns,
have to be considered.
Mainly caused by detector resolution, the time scale $\tau$ is
placed in the nanosecond regime.
The time scale for the nuclear-resonant response of sample and reference
is determined by the lifetime of the nuclear excited state, which
is 141\,ns for the 14.4125\,keV nuclear level of the \Fe\ isotope
used here.
%
%
\paragraph*{}
%
%
%
%
%
Experiments were performed at sector 3 of the
Advanced Photon Source.
The storage ring was filled with 23 electron bunches producing
about 6.2$\cdot$10$^6$ x-ray pulses per second.
The SR was monochromatized to about 2\,meV bandwidth to avoid
overload of the detector system while maintaining the spectral
brightness of the x rays around the nuclear transition energy.
After reduction of the beam size to 0.3$\times$0.3\,mm$^2$,
the photon flux was 10$^9$\,ph/s and about 3$\cdot$10$^7$\,ph/s
after the triple-Laue IFM.
A schematic of the experimental setup is shown in
Fig.\,\ref{fig:setup}.
Three situations were realized\,: the sample is placed
outside the IFM and magnetized collinear with the linear
polarization of the SR, as before but
magnetized perpendicular to polarization and propagation
direction of the SR,
the sample is mounted inside the IFM and magnetized parallel
to the polarization direction of the SR.
In all cases, the reference was a 1.1-$\mu$m-thick foil of
stainless steel (Fe$_{0.55}$Cr$_{0.25}$Ni$_{0.2}$) that was
95\,\% enriched in the \Fe\ isotope.
The sample outside the IFM was a 2.5-$\mu$m-thick iron foil
that was 53\,\% enriched in the \Fe\ isotope.
The control measurements with sample inside the IFM used a
2.2-$\mu$m-thick iron foil that was 95\,\% enriched in the \Fe\ isotope.
Very good knowledge of the time response of magnetized iron
and stainless steel motivated this choice.

%
%
%
%
\begin{figure}[t]
\centering
\includegraphics*[scale=0.55]{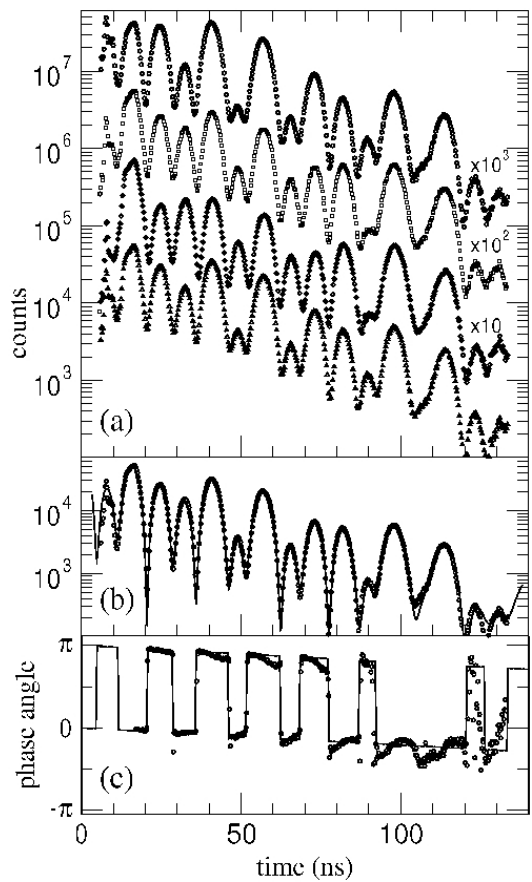}
\caption{Time spectra (a),(b) and phase angle (c) for the
control case, i.e., the sample mounted inside the IFM.
The time spectra (a) correspond to phase-retarder settings of
$\pi$, $-\pi/2$, 0, and $\pi/2$ (top to bottom).
The time spectrum (b) is the response of the sample only, i.e.,
the reference-beam path was blocked.
The solid line is a fit to the data.
The time-dependent phase angle was either extracted directly from
spectra (a) using Eq.\,(\ref{eq:phase1}), (symbols), or
calculated with parameters obtained from the fit of (b)
and an isomeric shift of 1$\Gamma$, (line).}
\label{fig:cntrl}
\end{figure}
For each experimental situation, we measured six time spectra\,:
the phase retarder adjusted to $\alpha=0,\pi,\pm\pi/2$, and
either IFM path blocked.
The spectra were collected in back-to-back sequences of 10\,min/spectrum
to minimize systematical errors caused by potential beam instabilities
and temperature drifts.
%
%
%
%
\begin{figure}[t]
\centering
\includegraphics*[scale=0.55]{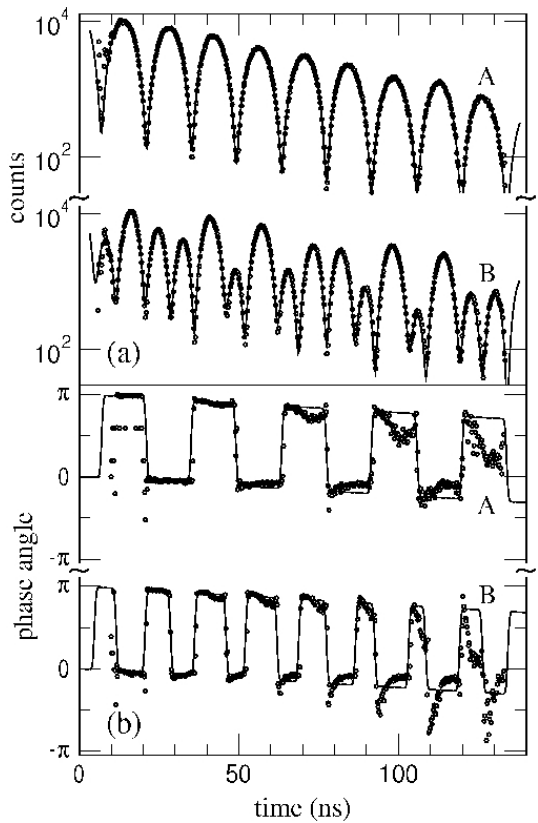}
\caption{Time-dependent intensities (a) and phase angles (b)
for the sample mounted outside the IFM and magnetized
perpendicular, A, or parallel, B, to the polarization of the SR.
The solid lines are fits to the intensity data.
The phase angles were either extracted directly from spectra
using Eq.\,(\ref{eq:phase1}) as explained in the text, (symbols),
or calculated with parameters obtained from the fits of (a)
and an isomeric shift of 1$\Gamma$, (line).}
\label{fig:IntPhase}
\end{figure}
Temperature variations during the experiment were less than 150\,mK.
For the control case, the relative phase change of the x-rays
by the sample is calculated from the four interference spectra $I(\alpha)$
without specific knowledge about the IFM imperfections.
With $\xi=I(0)-I(\pi)$ and $\eta=I(\pi/2)-I(-\pi/2)$, we obtain from
ref.\,\cite{Sturhahn01} for the time-dependent phase angle
\begin{equation}
\ee^{\ei\phi}=\frac{\xi+\ii\eta}{\sqrt{\xi^2+\eta^2}}\:\:.
\label{eq:phase1}
\end{equation}
The time spectra for the different phase-retarder settings, the
time spectrum of the sample alone, and the derived phase angle,
$\phi$, are shown in Fig.\,\ref{fig:cntrl}.
A standard evaluation code, {\sc Conuss} \cite{Sturhahn00}, was used
to fit the spectrum of the sample.
The excellent agreement supports the notion of well-defined
hyperfine interaction parameters of the chosen iron foil.

%
%
%
%
\begin{figure}[b]
\centering
\includegraphics*[scale=0.6]{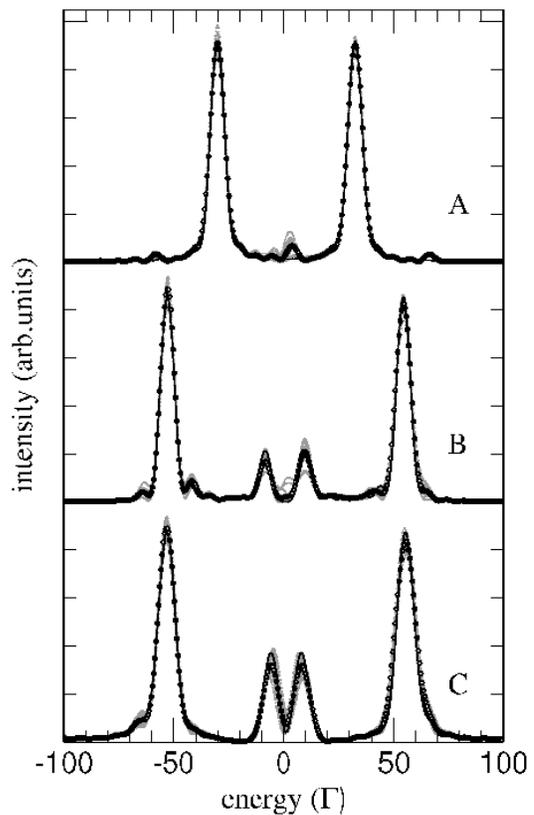}
\caption{Reconstructed energy spectra
for the exo-interferometric case and sample magnetized
perpendicular, A, or parallel, B, to the polarization of the SR,
and the control case with sample inside IFM and magnetized
parallel, C.
The solid lines are fits to the average spectra (circles).
Spectra from individual sequences are shown as shaded dots.
The energy scale is given in units of the nuclear level width
$\Gamma$=4.66\,neV.}
\label{fig:Recon}
\end{figure}
If the sample is placed outside the IFM, Eq.\,(\ref{eq:field})
is applied, and, if one neglects the convolution integral,
the relative phase results from Eq.\,(\ref{eq:phase1}) with the
replacement $\xi=I(0)-I(\pi)-4f\,I_S$.
The last term is composed of the time spectrum of the sample, $I_S$,
and a factor $f$ that accounts for absorption in the reference and
IFM imperfections.
$f$ is the product of IFM visibility, 0.45 in this case, and intensity
ratio of the individual IFM paths.
We determined experimentally $f=0.504$.
Fig.\,\ref{fig:IntPhase} shows the time-dependent intensities and
phases for the two directions of magnetization with the sample
outside the IFM.
In Figs.\,\ref{fig:cntrl} and \ref{fig:IntPhase}, one notices
overshoots at phase jumps, i.e., at sign changes of the transmitted
electric field associated with zero intensity.
In those minima, the measurement is most sensitive to effects of the
detector resolution and potentially other systematic errors.
However, the resulting phase errors are associated with very small
field amplitudes, which reduces their significance.
%
%
\paragraph*{}
The determination of the time-dependent phase as described
in the previous section did not require specific knowledge
of reference or sample.
The phase angle together with the measured intensities of
the sample alone, as displayed in Figs.\,\ref{fig:cntrl} and
\ref{fig:IntPhase}, permits us to reconstruct the energy
response $\tilde{S}(E)$ of the sample by Fourier transformation
\begin{equation}
\tilde{S}(E)=
\int \sqrt{I_S(t)}\,\ee^{\ei\phi(t)}
\ee^{\ei E t/\hbar}\,dt\:\:.
\label{eq:reconFT}
\end{equation}
In this expression, $I_S(t)$ is the time spectrum of the
sample, shown in Figs.\,\ref{fig:cntrl}(b) and \ref{fig:IntPhase}(a),
and $\phi(t)$ is the phase angle determined earlier and shown in
Figs.\,\ref{fig:cntrl}(c) and \ref{fig:IntPhase}(b).
$\tilde{S}(E)$ was calculated for each sequence of data (about 1h
collection time).
In Fig.\,\ref{fig:Recon}, we show the reconstructed energy
spectra $|\tilde{S}(E)|^2$ for each data sequence, the average
of all data sequences, and the calculated energy spectra.
The oscillation pattern in the time spectra and the separation
of the resonances in the energy spectra are dominated by the
magnetic field at the \Fe\ nuclei.
The results for magnetic field values agree within 0.03\,\%
for the control case and within 0.2\,\% and 0.7\,\% for the
exo-interferometric situations A and B.
In addition, the reconstructed energy spectra contain a small
but noticeable overall shift, the so-called isomeric shift.
One obtains 1.037(6)\,$\Gamma$ for the control case, in
agreement with the literature value of 0.9(2)\,$\Gamma$ \cite{MBdataIx74}.
The exo-interferometric situations A and B give 1.05(2)\,$\Gamma$
and 0.67(2)\,$\Gamma$ for the same quantity.
The values are given in units of $\Gamma$=4.66\,neV, the width of
the nuclear-excited state.
In our measurement, we used an avalanche photodiode detector with
about 1\,ns resolution, which results in an accessible energy range
in the reconstructed spectrum of about 140\,$\Gamma$=0.66\,$\mu$eV.
The energy resolution is related to the observed duration of the
time-delayed signal \cite{Sturhahn01}.
Here the time duration of 116\,ns was determined by the operating mode
of the Advanced Photon Source and detector dead time.
It leads to an energy resolution of 23\,neV.
The additional broadening of the resonances in Fig.\,\ref{fig:Recon}
is caused by thickness-related self-absorption effects in
the samples.
The agreement of calculation and reconstructed spectra is generally
very good even though Fig.\,\ref{fig:Recon} shows deviations, which
appear to be stronger for the exo-interferometric cases.
Also for the control case, a slight asymmetry in the spectrum can
be observed.
Possible explanations include inhomogeneity in sample or reference
and non random imperfections of the IFM.
At present, we have no conclusive explanation for these small but
noticeable effects.
In situation B, our neglect of the convolution term in Eq.\,(\ref{eq:field}),
which amounts to an approximation in the process of phase determination,
leads to larger deviations of, e.g., the measured isomeric shift.
A glance at Fig.\,\ref{fig:Recon}\,B shows that two resonances of the
sample are rather close to the origin of the energy scale, which marks
the position of the single resonance of the stainless-steel reference.
The overlap of the energy spectra leads to a stronger contribution
of the convolution term than in situation A, and its neglect
causes a larger distortion in the reconstructed energy spectra.

Exo-interferometry has been introduced as spectroscopic use of an
x-ray IFM with distinct advantages in sample utilization.
The ideas of exo-interferometry are general, but potential applications
need to use technically available time scales, which are essentially
given by the best time resolution of x-ray detectors.
At present, streak cameras provide ps resolution, which could make a
meV-energy range accessible with $\mu$eV resolution.
The presented data have unambigously demonstrated that time-dependent
phase changes in x-ray fields can be measured with samples positioned
inside or outside an x-ray IFM.
To demonstrate the principle, our studies were conducted in transmission
geometry.
But, in contrast to conventional x-ray interferometry, crystal diffraction
or surface reflection could be investigated as well.
For example, in pump-probe time-resolved x-ray diffraction
experiments \cite{Reis01}, exo-interferometric phase determination
may become a very useful tool.
It also becomes possible to manipulate the x-ray beam between IFM and
sample, e.g., by focusing or polarization control.

We want to thank J.\,Zhao for valuable support during beamline
operations.
This work was supported by the U.S. Department of Energy,
Basic Energy Sciences, Office of Science, under Contract 
No.\,W-31-109-Eng-38.
%
%
%
%

%
%
\end{document}